\documentclass{aastex63}

\definecolor{Red}{rgb}{1,0,0}

\usepackage{bm}

\received{} 
\revised{} 
\accepted{} 
\submitjournal{ApJ}

\shorttitle{Modeling ion beams observed by PSP}
\shortauthors{Ofman et al.}

\begin{document}

\title{Modeling ion beams, kinetic instabilities, and waves observed by the Parker Solar Probe near perihelia}

\correspondingauthor{Leon Ofman}
\email{ofman@cua.edu}

\author[0000-0003-0602-6693]{Leon Ofman}
\affiliation{Department of Physics\\
Catholic University of America  \\
Washington, DC 20064, USA}
\affiliation{Heliophysics Science Division\\
NASA Goddard Space Flight Center \\
Greenbelt, MD 20771, USA}

\author{Scott A Boardsen}
\affiliation{Partnership for Heliophysics and Space Environment Research\\
 University of Maryland\\
 Baltimore, MD 21250, USA}
\affiliation{Heliophysics Science Division\\
NASA Goddard Space Flight Center \\
Greenbelt, MD 20771, USA}

\author{Lan K Jian}
\affiliation{Heliophysics Science Division\\
NASA Goddard Space Flight Center \\
Greenbelt, MD 20771, USA}

\author[0000-0001-5030-6030]{Jaye L Verniero}
\affiliation{Heliophysics Science Division\\
NASA Goddard Space Flight Center \\
Greenbelt, MD 20771, USA}

\author[0000-0003-1138-652X]{Davin Larson}
\affiliation{Space Sciences Laboratory\\
University of California\\
Berkeley, CA 94720, USA}




\begin{abstract}
Recent in-situ observations from the Parker Solar Probe (PSP) mission in the inner heliosphere near perihelia show evidence of ion 
beams, temperature anisotropies, and kinetic wave activity, which are likely associated with kinetic heating and acceleration processes of the solar wind. In particular, the proton beams were detected by PSP/SPAN-I and related magnetic fluctuation spectra associated with ion-scale waves were observed by the FIELDS instrument. We present the ion velocity distribution functions (VDFs) from SPAN-I and the results of 2.5D and 3D hybrid-particle-in-cell (hybrid-PIC) models of proton and $\alpha$ particle super-Alfv\'{e}nic beams that drive ion kinetic instabilities and waves in the inner-heliospheric solar wind. We model the evolution of the ion VDFs with beams, ion relative drifts speeds, and ion temperature anisotropies for solar wind conditions near PSP perihelia. We calculate the partition of energies between the particles (ions) along and perpendicular to the magnetic field, as well as the evolution of magnetic energy and compare to observationally deduced values.  We conclude that the ion beam driven kinetic instabilities in the solar wind plasma near perihelia are important components in the cascade of energy from fluid to  kinetic scale, an important component in the solar wind plasma heating process.
\end{abstract}



\section{Introduction} \label{intro:sec}

It is well established that the dissipation of magnetized fluctuations in the solar wind occur at small kinetic scales below the ‘break point’ of the fluid-scale  MHD turbulent cascade \citep[see, e.g., recently][]{Mat20,Bow20a}  at frequencies near the proton gyro-resonance frequency. The ion kinetic instabilities in the inner heliospheric ($\leq1$ AU) solar wind plasma could be driven by ion temperature anisotropy and ion beams in multi-ion solar winds (SW) plasma, as deduced in the past from the analysis of spacecraft data such as Wind \citep[e.g.,][]{Kas08,Mar11,Mar12,Alt18,Alt19,Kas19}, Helios \citep[e.g.,][]{Mar82a,Mar82b,Hel13,Dur19}. Recent Parker Solar Probe observations substantiate this claim; for example, PSP Solar Wind Electrons Alphas \& Protons (SWEAP) \citep{Kas16} /Solar Probe Analyzers-Ions (SPAN-I) \citep{Liv21} VDFs and FIELDS \citep{Bal16}  electromagnetic waves instrument observations show evidence  that ion beams, kinetic instabilities and ion-scale waves play an important role in the dynamics of the SW plasma at never before explored perihelia regions \citep[e.g.][]{Ver20,Bow20a,Bow20b,Kle21,Vec21}. The ion kinetic instabilities as evident from the non-Maxwellian ion VDFs, and the wave dispersion and dissipation, provide evidence in support for the heating at the dissipation scale as expected from the turbulent energy cascade theory from MHD fluid to small (kinetic) scales \citep[e.g.,][]{Zan18,Zan21,Adi21}.

 In  coronal holes associated with the sources of the fast solar wind, the protons and ions are observed to be hotter than electrons and anisotropic, demonstrating the importance of anisotropic proton and heavier ion heating \citep[e.g.,][]{Koh97,CFK99,HS13,Jef18}. Moreover, $\alpha$ particles are typically hotter and faster than protons in the fast SW \citep[e.g.,][]{Mar82b,BG14,Dur19,Dur19b}  and could indicate the SW source regions \citep{Ofm04,Gio07,OK10,Abb16,Mos20}. The relative abundance of $\alpha$ particles in the solar wind varies considerably with the wind type and the solar cycle \citep[e.g.,][]{Kas07,Dur17,Alt19}. Analysis of spectroscopic observations shows that the heavy ion population may also  become hotter and anisotropic in coronal streamers \citep[e.g.][]{FCK03,Abb16,Cra20,Zha21} relevant to PSP measurements of the slow SW in the ecliptic plane.


In the present study we employ the well-established 2.5D and 3D hybrid modeling approach to study ion beams and related instabilities in electron-proton-$\alpha$ solar wind-like plasma developed by \citet{Ofm10} and \citet{Ofm19}{\bf,} expanding on past methods (see Section~\ref{model:sec} for details).  The ion instabilities, nonlinear wave-particle interactions, and kinetic processes associated with the heating of ions are well described by the hybrid modeling approach, and the numerical noise is reduced to orders of magnitude below the modeled physical process level with adequate number of numerical particles, resolution, and well-tested numerical noise reduction techniques. Naturally, the purpose of the hybrid models is to focus on the SW parameter space where electrons can still be treated as a fluid. The nonlinear hybrid modeling approach uses the Particle-In-Cell (PIC) method for the ions, while electrons are treated as a  background fluid through a generalized Ohms' law. The model enables investigation of the observed self-consistent non-equilibrium ion VDFs, ion-scale wave spectra and the nonlinear dispersion relation. The advantage of this approach over other methods such as full-PIC is the computational cost effectiveness that is crucial for computationally demanding 3D modeling, where only ion kinetic timescales need to be resolved.

The hybrid models were used recently to model related multi-ion plasma instabilities in the SW{\bf; for} example, the acceleration of protons and $\alpha$ particles by large amplitude Alfv\'{e}n waves in the SW plasma was demonstrated using 1.5D hybrid model by \citet{MVO13}. The effects of proton-$\alpha$ super-Alfv\'{e}nic drift in inhomogeneous expanding solar wind was studied recently using 2.5D \citep{OVM14,MOV15,OOV15,OVR17}, as well as turbulence in solar wind at ion scales \citep{RO19}. Several 3D hybrid models of SW plasma were recently developed and applied primarily to study turbulence cascade to ion scales in SW plasma \citep[e.g.][]{Vas15,Vas20,Cer17,Fra18,Hel19,Mar20}. Recently, the nonlinear evolution of the proton-$\alpha$ drift instability and the $\alpha$ particle temperature anisotropy driven instability were modeled using the full 3D hybrid model \citep{Ofm19} for the expected parameters close to the Sun ($\sim10R_s$), a region to be explored by PSP at perihelia (Encounters 22-24 in the years 2024-2025).

The present paper is the first multi-dimensional hybrid modeling nonlinear study of super-Alfv\'{e}nic proton and the $\alpha$ particle beams and related instabilities in SW plasma. The paper is organized as follows. In Section~\ref{obs:sec} we discuss the PSP/SPAN-I and FIELDS instrument observation near perihelia that provide the observational motivation for our numerical study. In Section~\ref{model:sec} we present the detail of the hybrid model. Section~\ref{num:sec} is devoted to the numerical results, and Section~\ref{disc:sec} contains the discussion and conclusions, including the implications for PSP observations near perihelia.

\section{Observational Motivations} \label{obs:sec}

\begin{figure}[h]
\centering
\includegraphics[width=\linewidth]{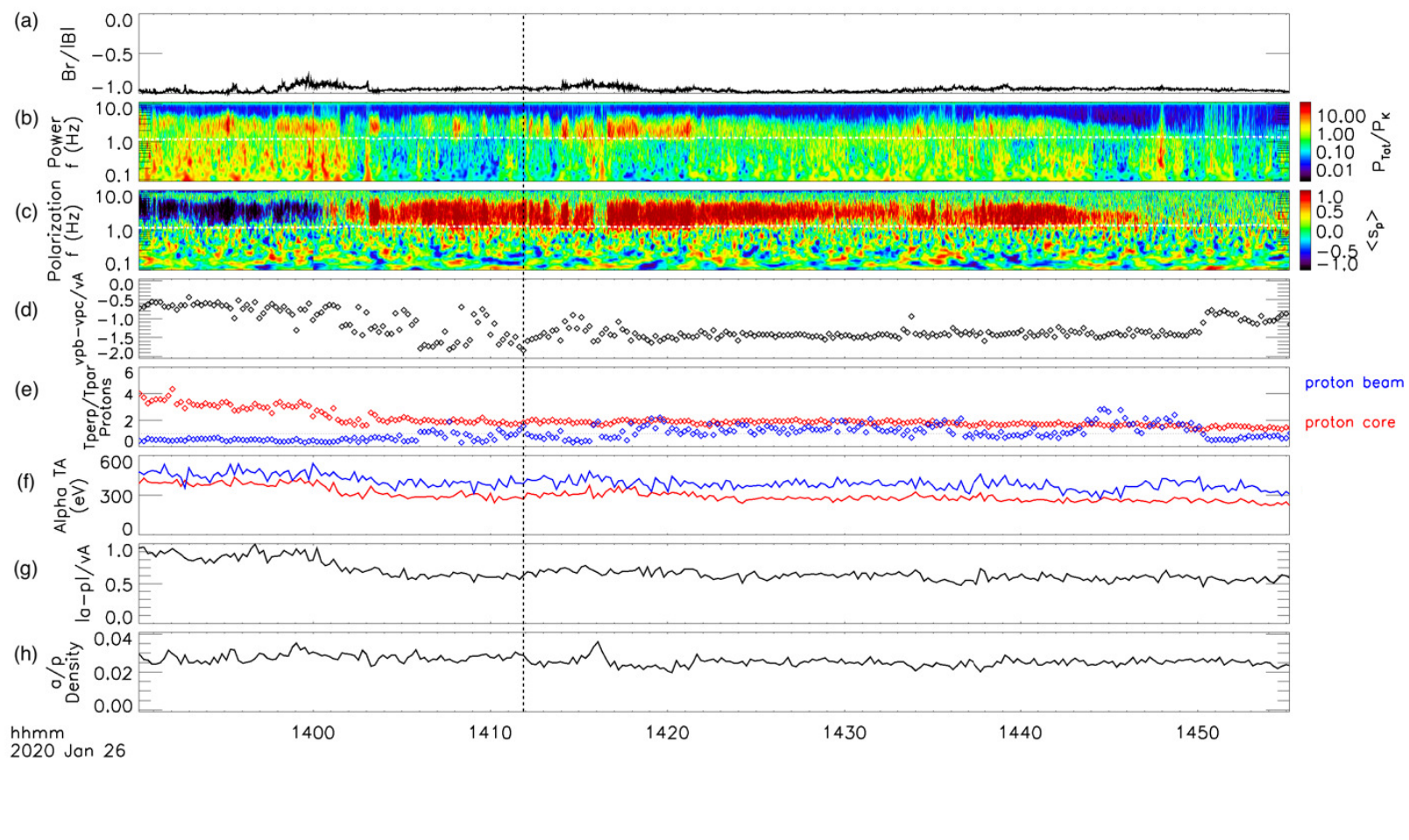}
\caption{Example wave event at 36$R_s$ featuring a change in polarization from LH (blue) to RH (red). Shown is the (a) ratio of radial magnetic field component to total field magnitude, (b) wave power of $\mbox{\bf B}$, (c) perpendicular polarization of $\mbox{\bf B}$, (the white dash lines in (b) and (c) indicate the proton resonant frequency at $\sim 1$ Hz) (d) proton fits of beam-core drift speed, normalized by $V_A$, (e) proton temperature anisotropy fits of the core (red) and beam (blue), (f) $\alpha$ particle parallel (red) and perpendicular (blue) temperatures, (g) $\alpha$-proton drift speed, normalized by $V_A$, and (h) $\alpha$-proton density ratio. }
\label{psp_fields01262020:fig}
\end{figure}

\begin{figure}[h]
\centering
\includegraphics[width=0.75\linewidth]{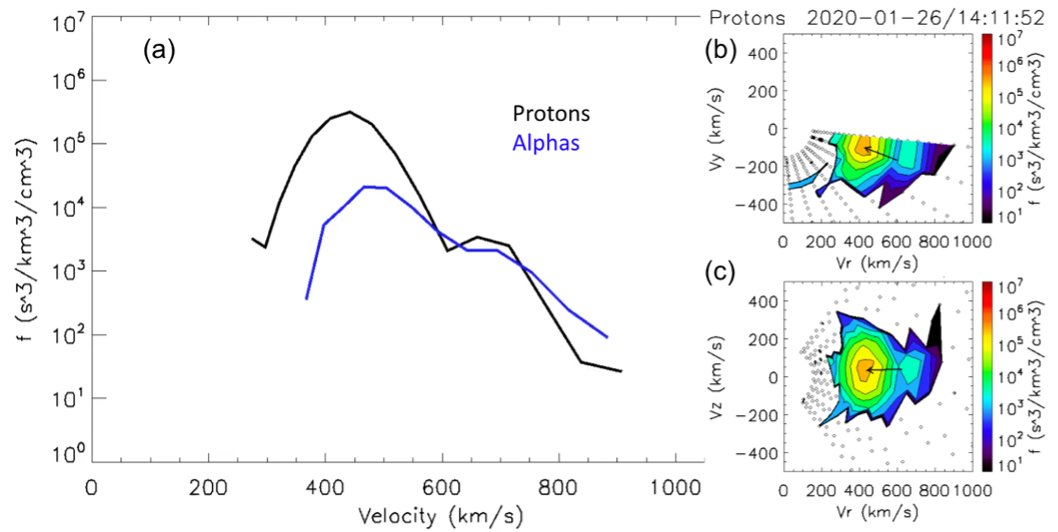}
\caption{Example proton and $\alpha$ VDFs during wave event at time indicated by dashed black line in Figure~\ref{psp_fields01262020:fig}. (a) VDF of protons (black) and $\alpha$ particles (blue), where all angles were summed and collapsed over energy. (b) Proton VDF contour elevation summed and collapsed onto azimuthal plane. (c) Proton VDF contour elevations summed and collapsed onto meridional  plane. The black arrow indicates the direction of the magnetic field, where the arrowhead is at the solar wind velocity and the length is $V_A$. }
\label{psp_vdf01262020:fig}
\end{figure}

The observational motivation for the present study is provided from PSP by the SPAN-I \citep{Liv21,Kas16} and FIELDS \citep{Bal16} instrument data. SPAN-I is a top-hat electrostatic analyzer that measures 3D ion velocity-space distributions and can separate protons from other heavy ions, such as $\alpha$ particles. Figure~\ref{psp_fields01262020:fig} shows an example wave event during PSP Encounter 4 (E4) at $\sim36R_s$ on 26-Jan-2020 for the time period 13:50-14:55 UT. Panel (a) shows that the magnetic field $\mbox{\bf B}$ was nearly quiet and radially aligned. Wavelet analysis revealed in (b) that in the spacecraft frame, there was coherent ion-scale wave activity at $\sim$3 Hz (slightly above the proton gyrofrequency at $\sim1$ Hz marked by the white dashed line), with a change in circular polarization signature from left-handed (blue) to right-handed (red) in panel (c). The waves appear at frequencies slightly larger than the proton gyro-frequency due to the (uncomputed) Doppler-shift from the plasma frame to the spacecraft frame. Note that the Doppler shift was not calculated due to previously discussed unresolved ambiguity \citep{Ver20,Bow20b}. The quantity $<S_p>$ indicates the degree of circular polarization, and $P_{tot}/P_k$ is the wave power normalized by background magnetic field fluctuations power. To extract wave power and polarization properties, we employ the Morlet wavelet transform to the FIELDS magnetometer data \citep{Ver20,Bow20a,TC98}. Double bi-Maxwellian fits in (d) of the proton VDFs reveal that the beam-core drift speed (normalized by the Alfv\'{e}n speed, $V_A$) may be regulating the wave-particle energy transfer process, as well as the proton temperature anisotropy of the core (red) and beam (blue) in (e). Other quantities that appear to qualitatively correlate with wave features are the preliminary plasma parameters obtained by the SPAN-I $\alpha$ particle moments, such as (f) $\alpha$-proton parallel (blue) and perpendicular (red) temperatures, (g) $\alpha$-proton drift speed normalized by $V_A$, and (h) $\alpha$-proton density ratio. To mitigate error due to SPAN-I partial field of view (FOV) obstruction by the heat shield, the $\alpha$ temperature anisotropy in Figure~\ref{psp_fields01262020:fig}f was obtained by transforming the $\alpha$ particle temperature tensor to field-aligned-coordinates and extracting the perpendicular and parallel components. Here the overall $\alpha$ particle population produces $T_\parallel>T_\perp$, indicating a secondary $\alpha$ particle beam is present, as can be seen by the blue 1D VDF in Figure~\ref{psp_vdf01262020:fig}.

From the SWEAP-I data it was found that most of the time $T_\perp/T_\parallel >1$ using bi-Maxwellian fits, as evident in Figure~\ref{psp_fields01262020:fig}e. In the analysis of  the VDF  data, where the beam is subject to the beam instability and velocity space diffusion, the ion core and beam populations cannot be fully separated and evidently the fits provide larger $T_\parallel$ (than $T_\perp$) of the combined ion populations. In the hybrid model discussed below we model initially isotropic beam and core populations, and our model allows separating the populations unambiguously, following the time-dependent evolution of the beam and core VDFs and calculating the temperature  anisotropies unambiguously using moments of the VDFs throughout the simulations.

Figure~\ref{psp_vdf01262020:fig} shows an  example of proton and $\alpha$ particle VDFs at the time indicated by the dashed black line in Figure~\ref{psp_fields01262020:fig}. Shown in \ref{psp_vdf01262020:fig}a are the 1D VDFs of protons (black) and $\alpha$ particles  (blue) where the protons are exhibiting evidence of a bump-on-tail VDF. Panel \ref{psp_vdf01262020:fig}b shows proton VDF contours in the azimuthal plane of the instrument, showing that the core of the distribution is sufficiently in the SPAN-I FOV. Panel \ref{psp_vdf01262020:fig}c shows the VDF contours in the meridional plane, where we see two resolvable peaks in velocity space. Additional examples of proton beam VDFs at Encounters 4-8 can be found in \citet{Ver21}, that show strong perpendicular spreading of the proton beam VDFs.


\section{Hybrid-PIC Model, Boundary, and Initial Conditions} \label{model:sec}

We use our recently developed parallelized 2.5D and 3D hybrid codes, based on the same principles as the 1D hybrid code originally developed by \citet{WO93}, and later expanded to a 2.5D hybrid code \citep{OV07} and parallelized by  \citet{Ofm10} to model the proton-$\alpha$, magnetized SW plasma, recently extended to full 3D hybrid model \citep{Ofm19} (see the above references for details of the method of solution and the normalization). Here we mention briefly that the velocities are normalized by the proton Alfv\'{e}n speed, $V_A$, the distances by the proton inertial length $\delta_p=c/\omega_{pp}$, where $\omega_{pp}$ is the proton plasma frequency), and the time is normalized by the inverse proton gyroresonat frequency $\Omega_p^{-1}$, where  $\Omega_p=eB/m_pc$, where $B$ is the background magnetic field strength with $\mbox{\bf B}=B\mbox{\bf \^x}$, $m_p$ is the proton mass, and $c$ is the speed of light. Quasi-neutrality of the plasma is assumed, i.e., $n_p+2n_\alpha=n_e$, where $n_p$ is the proton number density, $n_\alpha$ is the $\alpha$ particle number density, and $n_e$ is the electron number density. 

Ion kinetic dissipation scale is determined by the thermal proton gyroradius $r_p=V_{th,p}/\Omega_p$ or the ion inertial length $\delta_p$. These scales are typically of order $\lesssim$10 km in the solar wind at the distance of 36$R_s$ from the Sun using the measurements from PSP at E4 \citep[e.g.,][]{Ver21}. For example, the proton density at E4 was $\sim$1000 cm$^{-3}$, resulting in $\delta_p\sim7$ km, while for proton temperature of 20 eV the thermal proton gyroradius is $r_p\sim 4$ km. 
The kinetic wave dissipation wavelength scale is typically an order of magnitude larger than $\delta_p$, as evident from the dispersion relation. The proton gyration is well resolved with the time step $0.02\Omega_p^{-1}$, and a `thermal' proton travels $\sim0.1\delta_p$ in one time step. The electron fluid equation in the model, i.e., the generalized Ohm’s law has the capability to include the resistive term. However, no resistivity was used in present model. Test with normalized empirical resistivity coefficient of $\eta=0.001$ did not affect the results.

\begin{figure}[h]
\begin{center}
\includegraphics[width=0.5\linewidth]{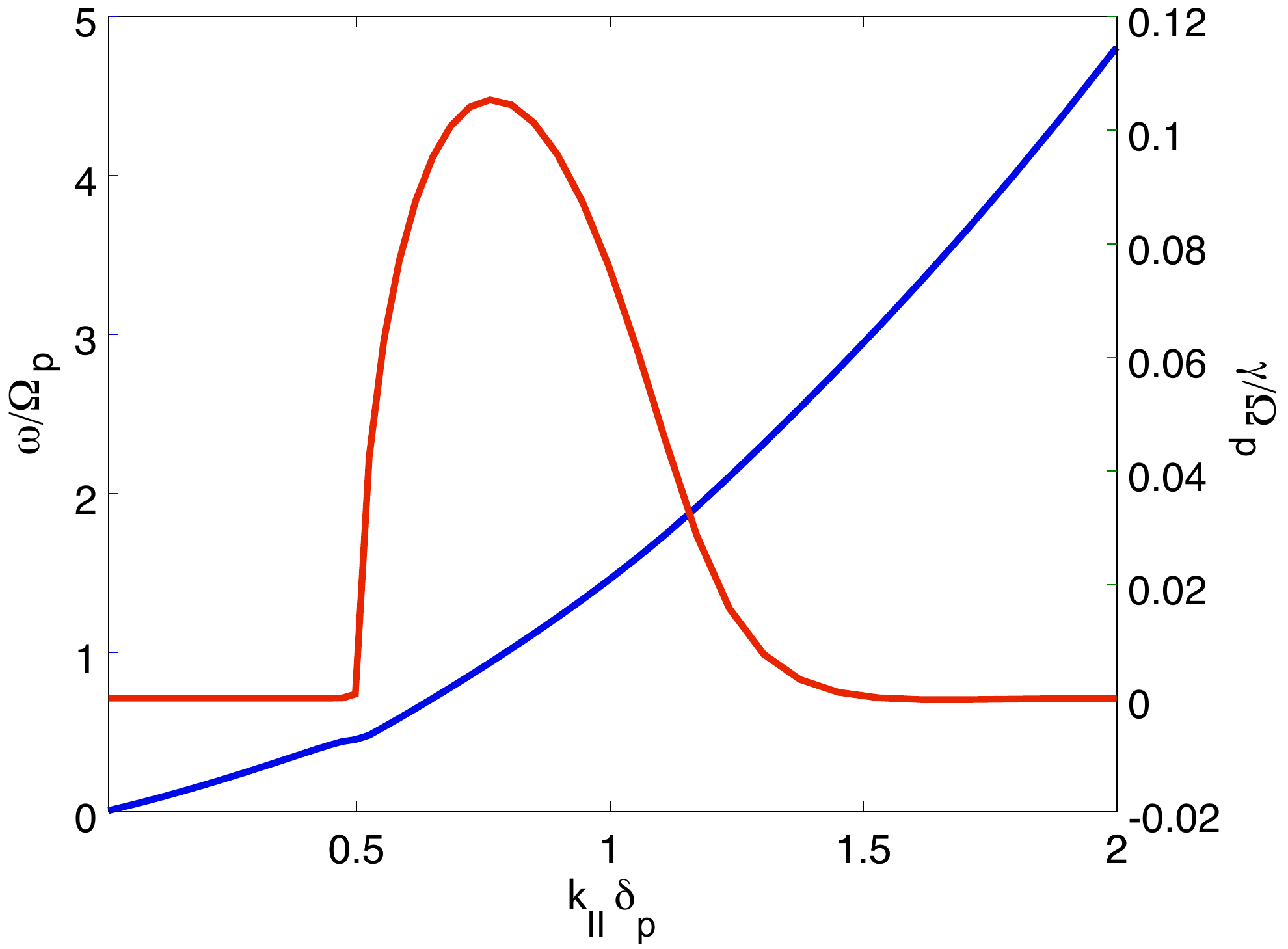}
\end{center}
\caption{The linear Vlasov's dispersion relation for the proton beam parameters in Case~3 (beam velocity magnitudes $V_{b,p}=2V_A$). The normalized real frequency $\omega$ (blue), and the growth rate $\gamma$ (red) dependence on $k_\parallel$ are shown. The fastest growing mode with $\gamma=0.105\Omega_p^{-1}$ is at $k_\parallel=0.76$.}
\label{displin:fig}
\end{figure}

The 2.5D hybrid codes solves three components of about $3.4\times10^7$ particle velocities that are  used to calculate the currents and the fields on the 2D spatial grid of $256\times256$ cells with 512 particles per cell (ppc) and grid size 0.75$\delta_p$. Thus, the size of the computational domain is $(192\delta_p)^2$. In the present study the 3D hybrid code uses $128^3$ cells with up to 128 ppc and grid size 1.5$\delta_p$ providing the 3D domain of $(192\delta_p)^3$. We note that each numerical super-particle is modeled by a Gaussian charge distribution that is spread over several cells and represents many  physical particles. In warm multi-ion plasma, the shortest length-scales to be resolved are determined by the ion kinetic dissipation scale. The solution of the Vlasov linear dispersion relation \citep[e.g.,][]{XOV04,OV07,MOV15,Ver20} shows that the proton and ion cyclotron waves are heavily damped at normalized $|k_\parallel|\gtrsim 1$ (normalized by $\delta_p^{-1}$). It is also evident from the linear Vlasov's dispersion relation analysis \citep[see, e.g., ][Figure 9]{Sti92,Ver20} that the fastest growing mode is well resolved. In Figure~\ref{displin:fig} we show the solution of the linear Vlasov's dispersion relation for the proton beam parameters of Case~3 (see Table~\ref{param:tab}) for electron-proton plasma and parallel propagation. It is evident that the fastest growing mode is at $k_{\parallel,\gamma_{max}}=0.76$, and the corresponding growth rate is $\gamma_{max}=0.105\Omega_p^{-1}$. Thus, the linear growth stage of the instability is on the order of tens of gyroperiods. Increasing the beam relative density results in faster growth rate with somewhat shorter $k_{\parallel,\gamma_{max}}$ in the linear dispersion.

Thus, our code resolves well the corresponding length scales down to the dissipation range and below, and the largest scale is more than an order of magnitude larger than the dissipation scale. The required number of particles per cell is determined by the limitation on the overall statistical noise, and the velocity space resolution which is typically a function of ion $\beta_i$. A numerical convergence test and total energy conservation monitoring were used to determine that the noise level is below the physical fluctuations level in the hybrid codes.  The equation of the hybrid model and the details of the method of solution can be found in our past studies \citep{OV07,OVM11,Ofm19}. In the present study we employ periodic boundary conditions in an initially homogeneous $e-p-\alpha$ plasma. The hybrid model computes the self-consistent evolution of the ion VDFs and includes the nonlinear effects of wave-particle interactions without further simplifying assumptions and is well suited to describe the nonlinear saturated state and relaxation of the magnetized plasma ion kinetic instabilities. In the present study we have used the expanding box model \citep{GV96,LVG01} to model the solar wind gradual expansion with the expansion rate parameter $\epsilon=10^{-4}$, consistent with our previous studies \citep[see, e.g.,][for the expanding box model description, implemented in our 2.5D and 3D hybrid codes]{OVM11,OVR17,Ofm19}.

A self-consistent kinetic wave spectrum, produced by initially unstable (i.e., due to super-Alfv\'{e}nic beams) plasma VDFs, is produced by the hybrid model (see  below).  Initial Maxwellian drifting ion populations lead to the drift instability for super-Alfv\'{e}nic drift, resulting in non-Maxwellian VDFs in the nonlinear relaxed state in the nearly collisionless SW plasma.  In Table~\ref{param:tab} we list the various plasma parameters of the modeled cases in the present study.

\begin{table}[h]
\caption{Plasma parameters of the numerical model runs for the various cases in the present study. The initial beam-core drift speed  $V_{d0,p}$ for protons, the initial beam-core drift speed for $\alpha$ particles $V_{d0,\alpha}$, the initial proton core density $n_{p,c}$, the initial proton beam density $n_{p,b}$, the initial $\alpha$ core density $n_{\alpha,c}$, the initial $\alpha$ beam density $n_{\alpha,b}$, the $\beta$'s of the ion populations (defined in terms of $n_e$).}
\centering
\begin{tabular}{ccccccccccc}
\hline
Case \# & $V_{d0,p}$  & $V_{d0,\alpha}$  & $n_{p,c}$  & $n_{p,b}$ &  $n_{\alpha ,c}$ & $n_{\alpha,b}$ & $\beta_{p,c}$ & $\beta_{p,b}$ & $\beta_{\alpha,c}$ & $\beta_{\alpha,b}$\\ \hline
	1  &  2            &  2       &  0.819 &  0.091 &  0.03 & 0.015 &  0.429 & 0.0916 & 0.429 & 0.0916\\
	2  &  2            &  2       &  0.728&  0.182 &  0.03 & 0.015 &  0.429 & 0.0916 & 0.429 & 0.0916\\
    	3  &  2            &    0    &  0.9 &  0.1 & 0.045  & -  &  0.429 & 0.0916 & 0.429  &  -    \\
    	4 & 2.5          &     0     &  0.9  & 0.1 & - & -   & 0.429 & 0.0916 & -  &  -   
\end{tabular}
\label{param:tab}
\end{table}

\begin{figure}[h]
\begin{center}
\includegraphics[width=0.7\linewidth]{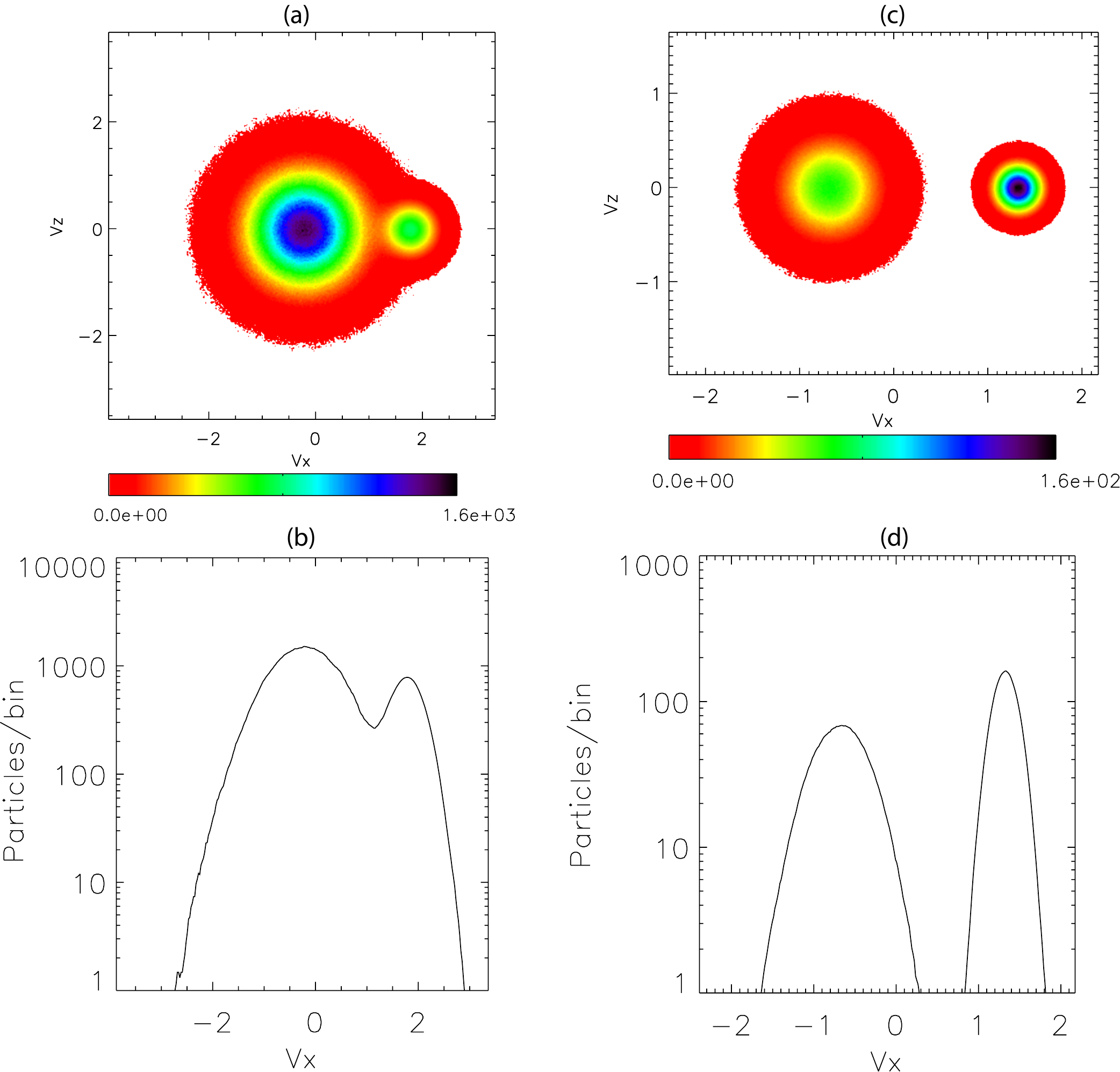}
\end{center}
\caption{The initial state of the ion VDFs for Case~1 with beam velocity magnitudes $V_{b,p}=V_{b,\alpha}=2V_A$. The core populations are on the left and the beam populations are on the right in the center of mass frames. (a) The proton VDF in $V_x-V_z$ phase space plane.  (b) The cut of the VDF along $V_x$ through the peak of the VDF shown in (a). (c) The $\alpha$ particle VDF in $V_x-V_z$ phase space plane. (d) The cut of the VDF along $V_x$ through the peak of the VDF shown in (c).}
\label{vxvz_pv2av2_128x3_iden0eps1e-4_dx0.75_t0:fig}
\end{figure}

In Figure~\ref{vxvz_pv2av2_128x3_iden0eps1e-4_dx0.75_t0:fig} we show the initial state of the VDFs in the $V_x-V_z$ phase space plane for protons (Figure~\ref{vxvz_pv2av2_128x3_iden0eps1e-4_dx0.75_t0:fig}a-b) and $\alpha$ particles (Figure~\ref{vxvz_pv2av2_128x3_iden0eps1e-4_dx0.75_t0:fig}c-d) for Case~1. The cut along $V_x$ through the peak of the VDFs is shown in the lower panels. In all cases the proton and $\alpha$ particle beams are lower density and lower $\beta_i$ compared to the core ion populations, in qualitative agreement with examples of PSP SPAN-I observations. The super-Alfv\'{e}nic drift speed between the beam and core is evident from the location of the peaks, the perpendicular temperatures and the relative number densities of the cores and beams are also evident from plots (1D cuts) of the 3D VDFs.

\newpage\section{Numerical Results} \label{num:sec}
In this section we present the results of the 2.5D and 3D hybrid modeling of the proton and $\alpha$ particle beams obtained from the 2.5D hybrid model, with parameters guided by PSP SPAN-I data. We show the evolutions of the ion VDFs, and the moments as evident in temperature anisotropies, parallel and perpendicular energies, and drift velocities of the various ion populations, as well as the associated magnetic energy and dispersions of the perpendicular magnetic fluctuations.
\subsection{2.5D hybrid modeling results}

\begin{figure}[h]
\centering
\includegraphics[width=\linewidth]{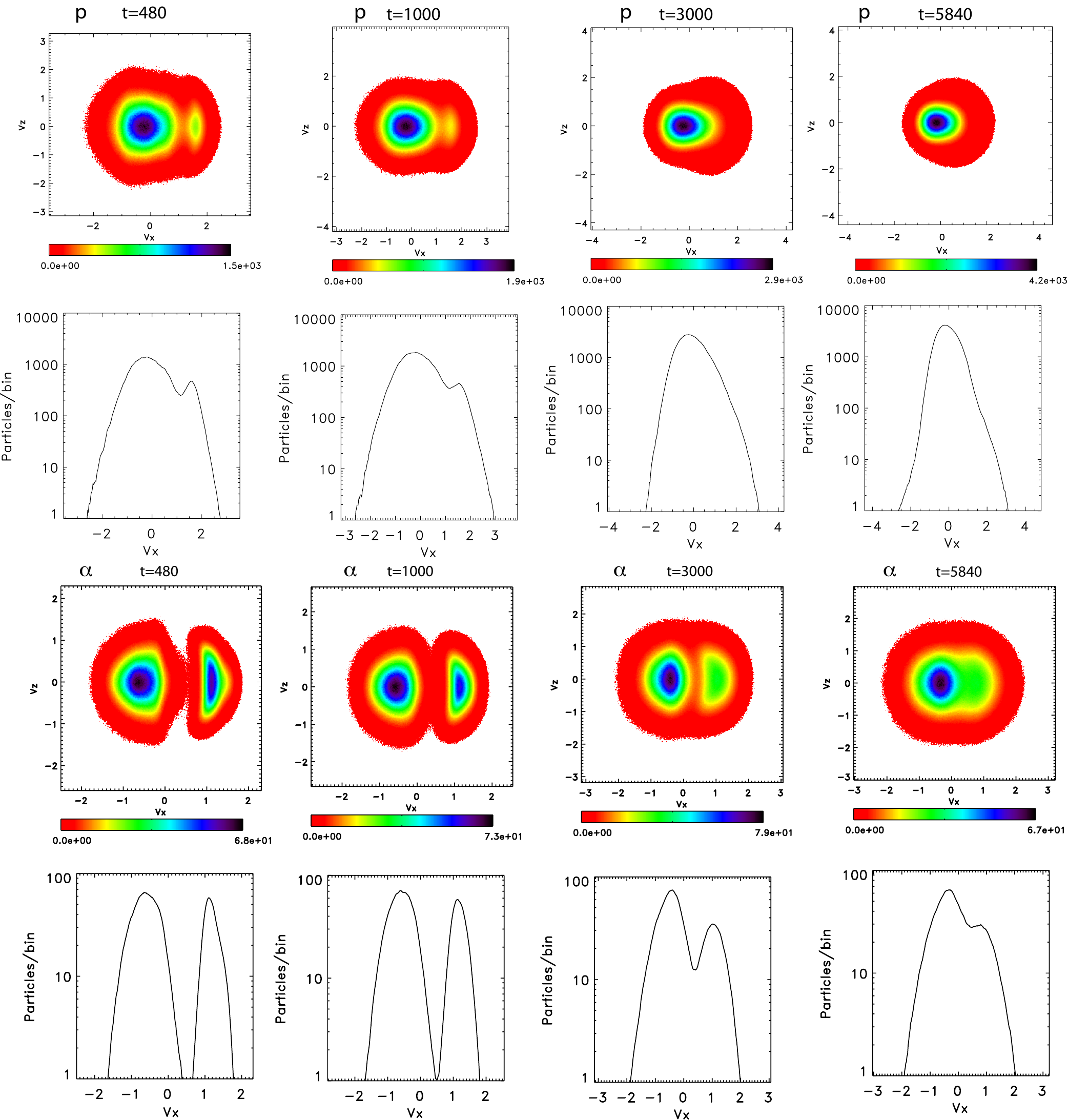}
\caption{Top two panels: the proton VDFs for Case~1 in the $V_x-V_z$ phase space plane at $t=480$, $t=1000$,  $t=3000$, and  $t=5840$ $\Omega_p^{-1}$. Lower panels: the cuts of the proton VDFs along $V_x$ through the peaks of the distributions at the corresponding times. Lower two panels: the $\alpha$ particle VDFs for Case~1 in the $V_x-V_z$ phase space plane at $t=480$, $t=1000$,  $t=3000$, and  $t=5840$ $\Omega_p^{-1}$. The cuts of the $\alpha$ particle VDFs along $V_x$ through the peaks of the distributions at the corresponding times.}
\label{vxvz_pv2av2_128x3_256_iden0eps1e-4_dx0.75_pa_t480-5840:fig}
\end{figure}

\begin{figure}[h]
\centering
\includegraphics[width=\linewidth]{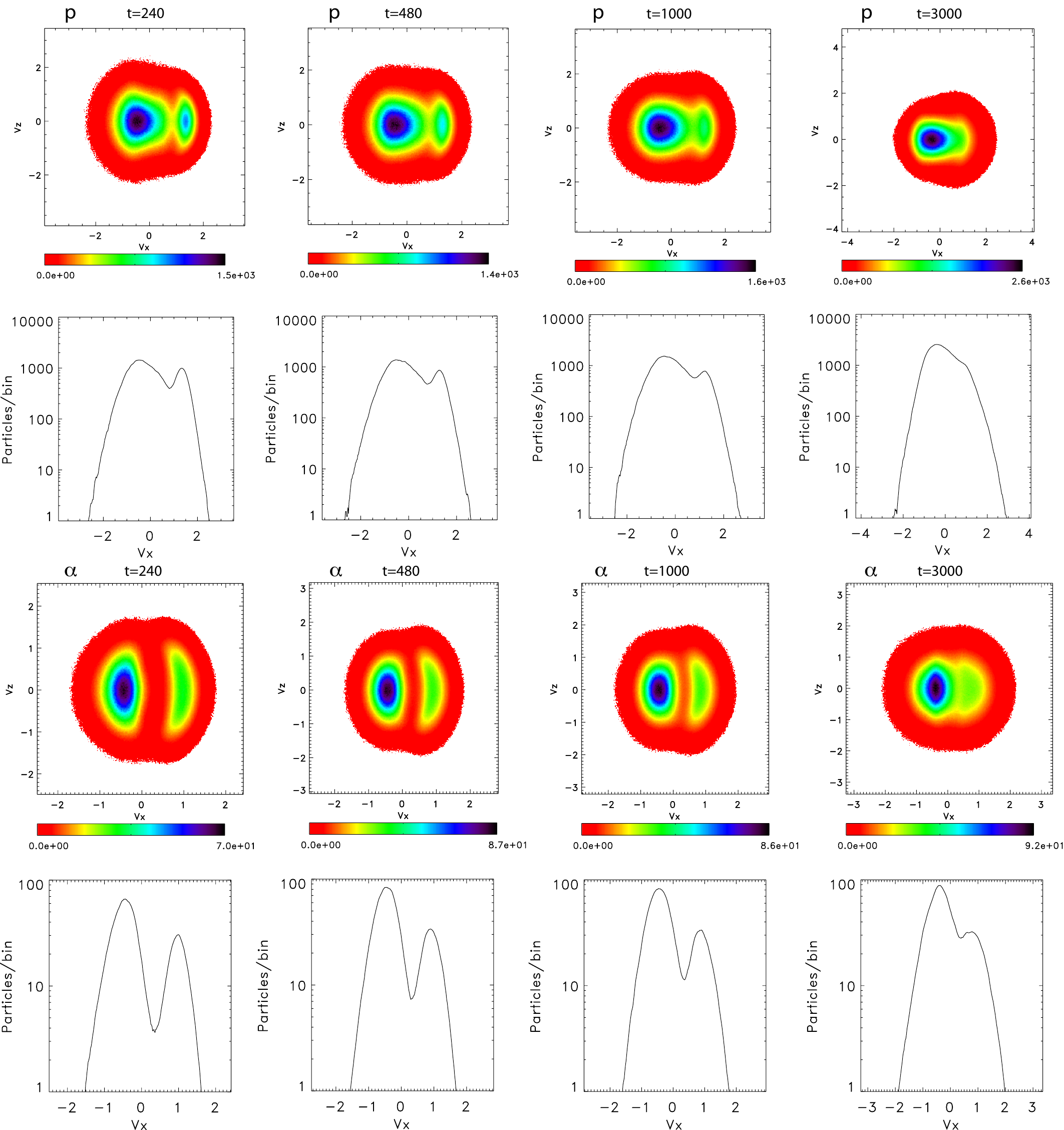}
\caption{Top two panels: the proton VDFs for Case~2 in the $V_x-V_z$ phase space plane at $t=240$, $t=480$,  $t=1000$, and  $t=3000$ $\Omega_p^{-1}$. Lower panels: the cuts of the proton VDFs along $V_x$ through the peaks of the distributions at the corresponding times. Lower two panels: the $\alpha$ particle VDFs for Case~2 in the $V_x-V_z$ phase space plane at $t=240$, $t=480$,  $t=1000$, and  $t=3000$ $\Omega_p^{-1}$. The cuts of the $\alpha$ particle VDFs along $V_x$ through the peaks of the distributions at the corresponding times.}
\label{vxvz_pv2av2npb0.182_128x4_256_iden0eps1e-4_dx0.75_pa_t240-3000:fig}
\end{figure}

\begin{figure}[h]
\centering
\includegraphics[width=\linewidth]{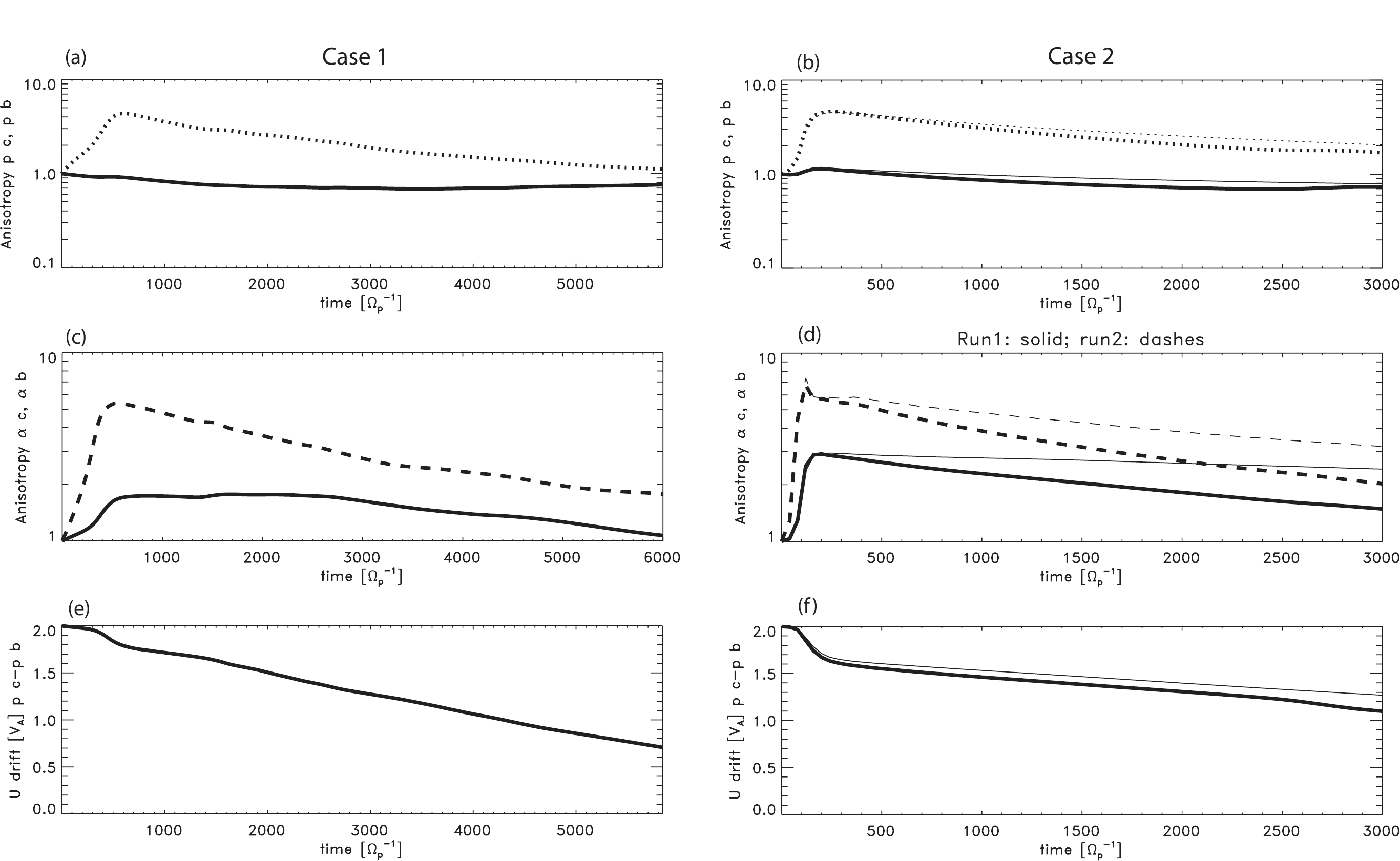}
\caption{The temporal evolution of the temperature anisotropies and the proton beam-core drift speed $U_{drift}$ with initial drift speed $V_d=2V_A$ for Cases~1 and 2. The temperature anisotropies of the proton core (solid) and beam (dashes) populations for (a) Case~1 and (b) Case~2. The $\alpha$ particle core (solid) and beam (dashes) populations temperature anisotropies for (c) Case~1 and (d) Case~2. The $U_{drift}$ for (e) Case~1 and (f) Case~2. The corresponding quantities calculated without solar wind expansion are added for comparison for Case~2 (thin line styles).}
\label{aniso_pv2av2_128x3_256_iden0eps1e-4_dx0.75_12:fig}
\end{figure}

\begin{figure}[h]
\centering
\includegraphics[width=\linewidth]{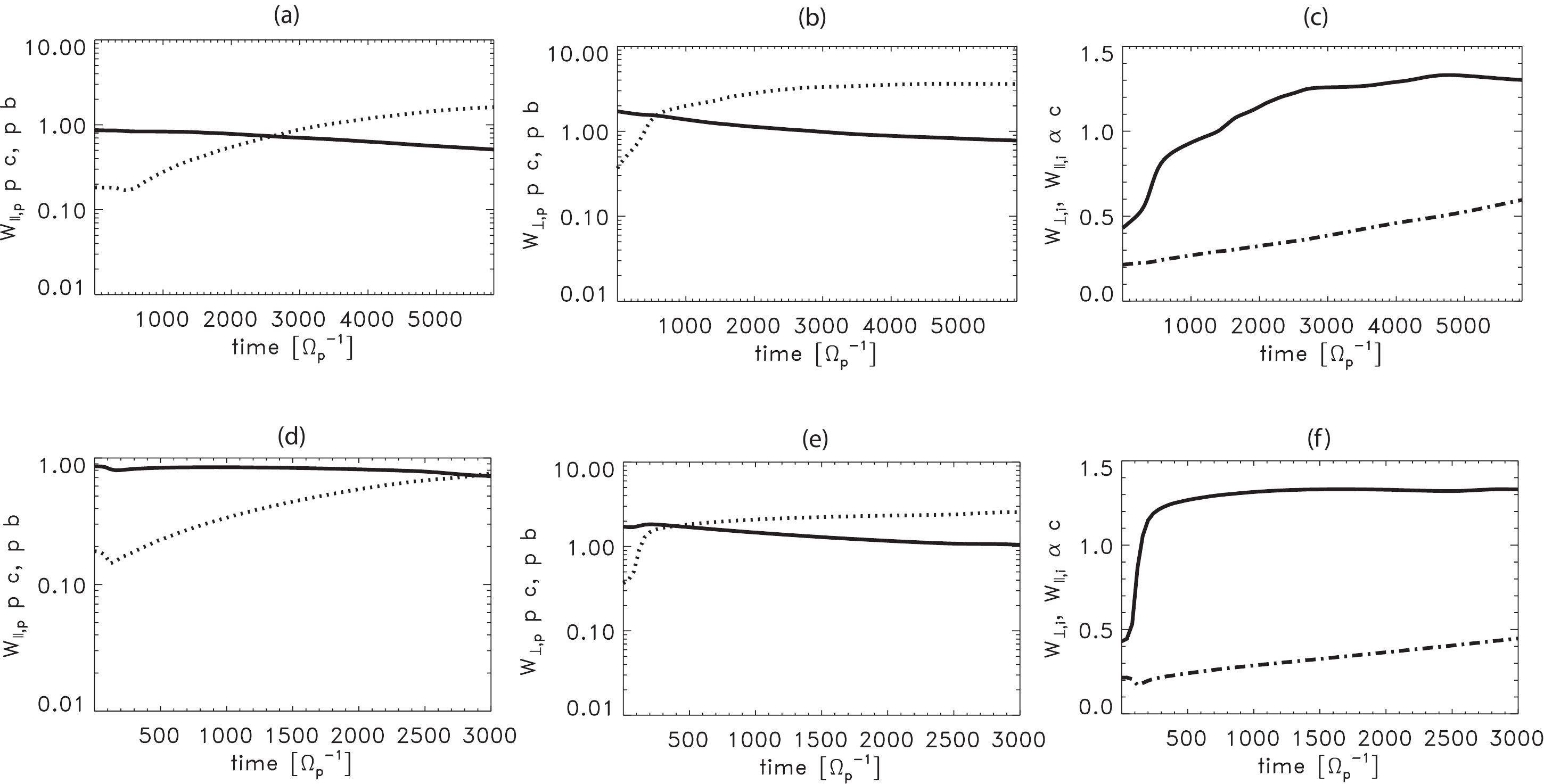}
\caption{The temporal evolution of the parallel and perpendicular energies for Case~1 of (a) $W_{p,\parallel}$ of the proton core (solid) and beam (dashes) populations, (b) $W_{p,\perp}$ of the proton core (solid) and beam (dashes) population (c) $W_{\alpha,\perp}$ (solid) and $W_{\alpha,\parallel}$ (dot-dashes) of the $\alpha$ particle core population. (d)-(f) same as (a)-(c) for Case~2.}
\label{wp_pv2av2_128x3_256_iden0eps1e-4_dx0.75_12:fig}
\end{figure}

\begin{figure}[h]
\includegraphics[width=\linewidth]{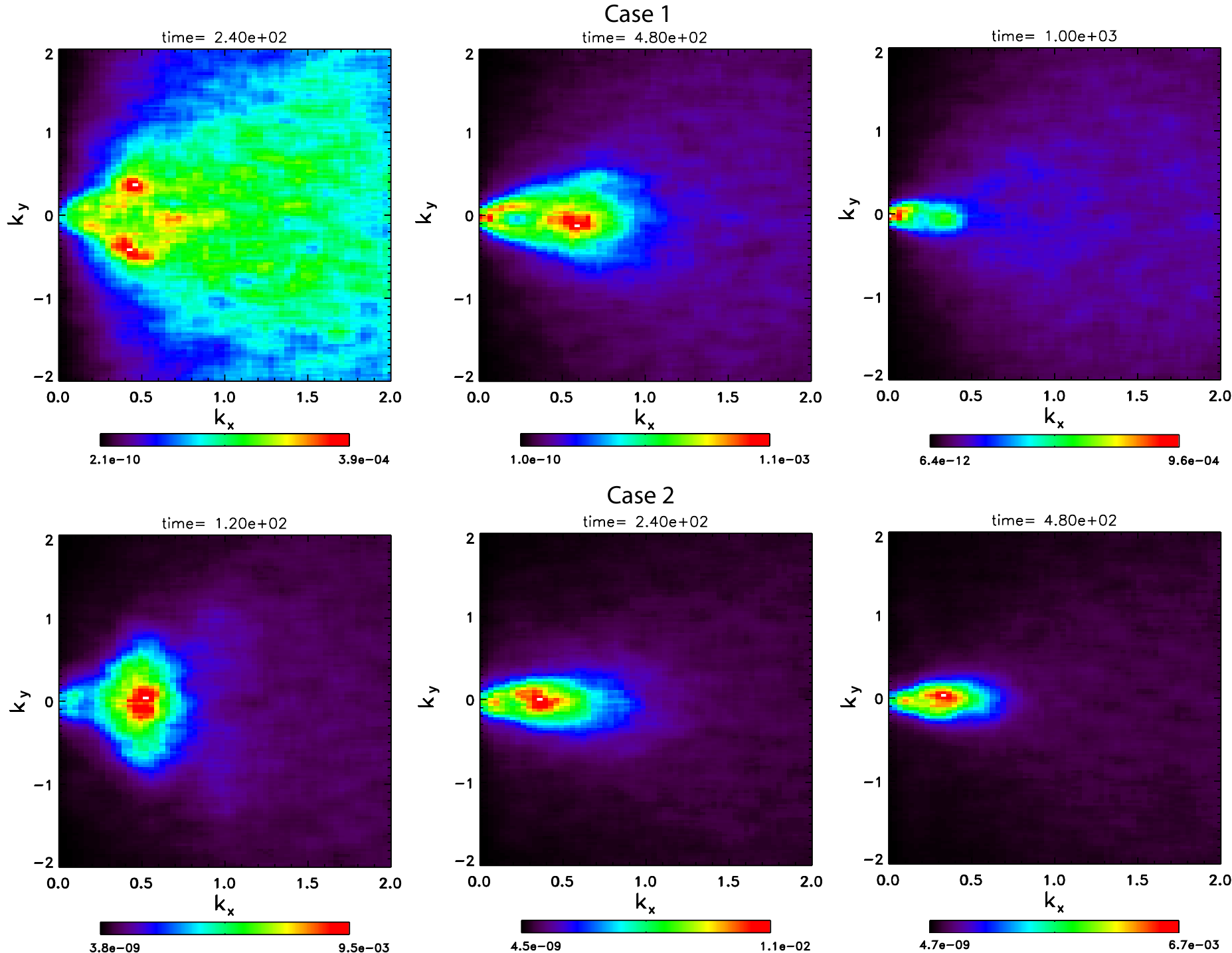}
\caption{The $k$-plane power spectra of the perpendicular magnetic fluctuations due to the beam instabilities modeled in Cases~1 and 2 with the 2.5D hybrid code. The wavenumbers $k_x$, $k_y$ are  normalized in units of $\delta_p^{-1}$. Top panels: Case 1 at $t=240$, 480, 1000 $\Omega_p^{-1}$. Lower panels: Case 2 at $t=120$, 240, 480 $\Omega_p^{-1}$.}
\label{2dfft_kxky:fig}
\end{figure}

\begin{figure}[h]
\includegraphics[width=\linewidth]{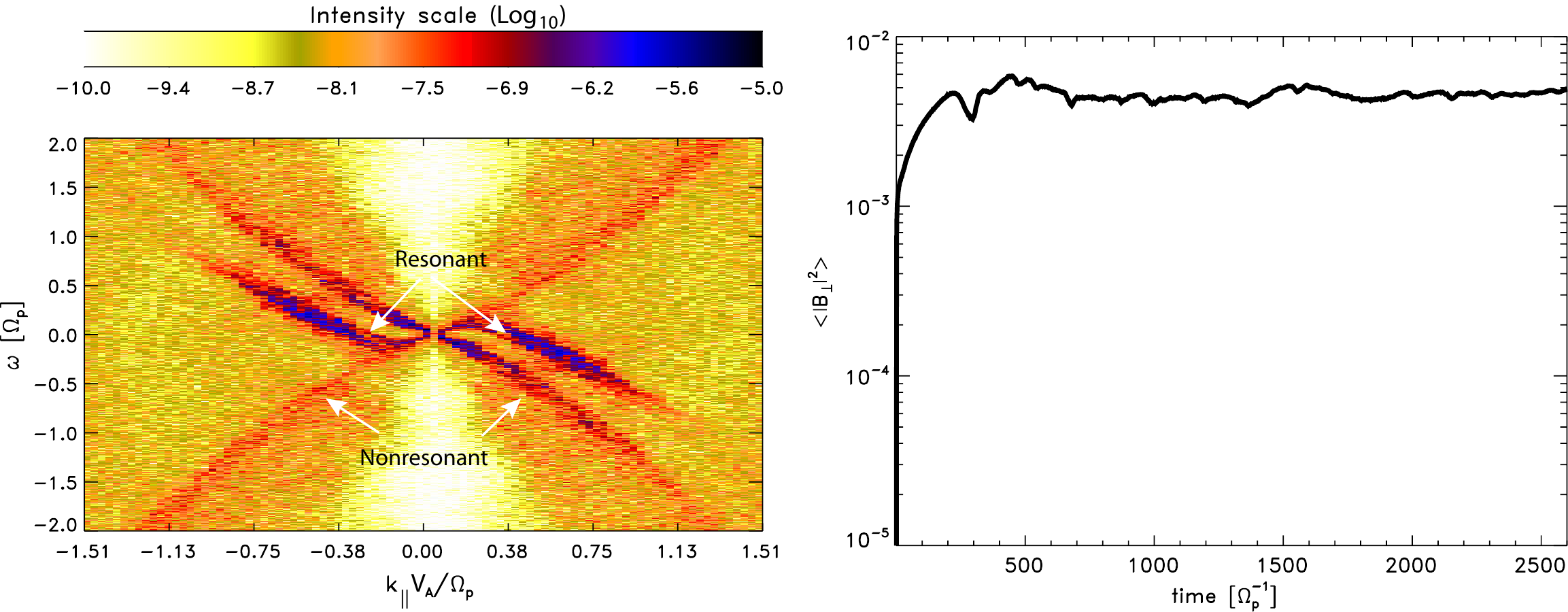}
\caption{Left panel: The dispersion relation calculated from the 2.5D hybrid model of the Fourier-transformed spatial-temporal fluctuations of $B_\perp$ in Case~1 using the time interval $t=0-1024\ \Omega_p^{-1}$. The white arrows identify the resonant and noresonant branches. The intensity scale is logarithmic $(Log_{10})$}. Right panel: the temporal evolution of the integrated $|B_\perp|^2$ averaged over the computational domain, i.e., the normalized perpendicular magnetic energy density.
\label{disp_pv2av2_128x3_256_iden0eps1e-4_dx0.75:fig}
\end{figure}

\begin{figure}[h]
\includegraphics[width=\linewidth]{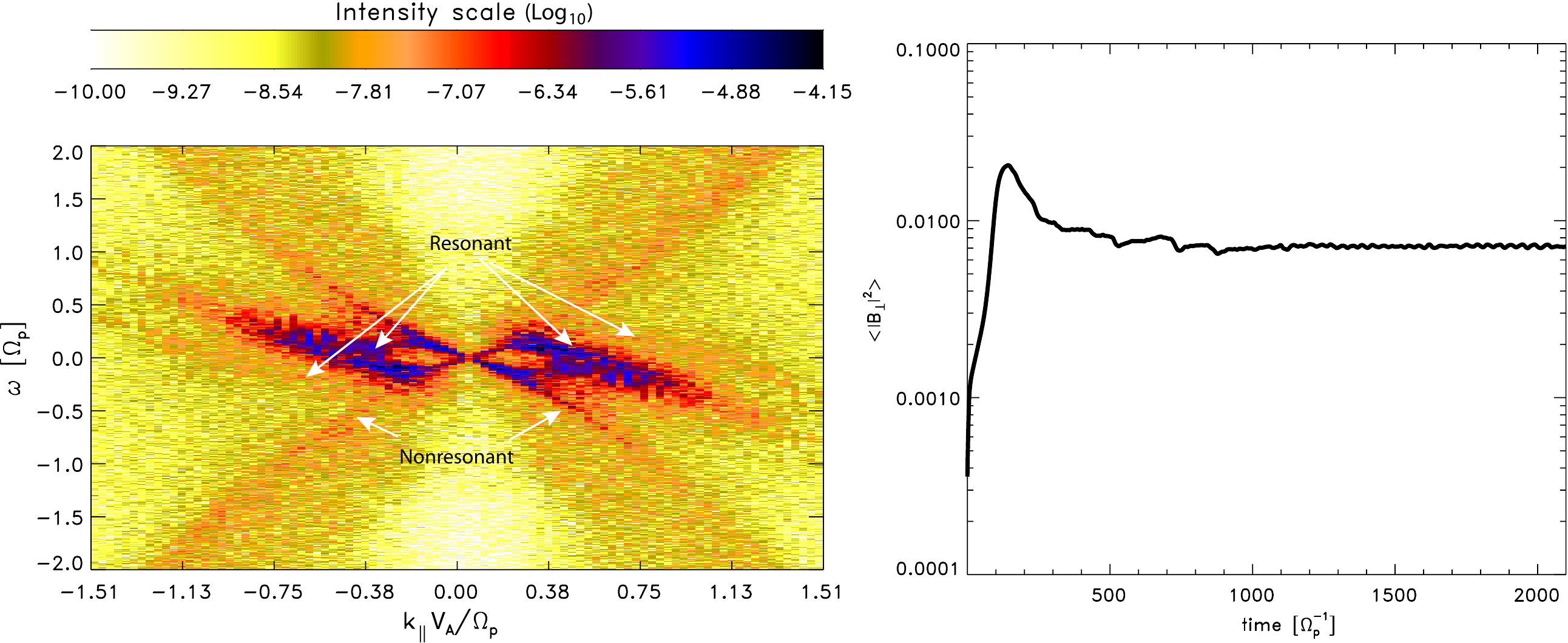}
\caption{Same as Figure~\ref{disp_pv2av2_128x3_256_iden0eps1e-4_dx0.75:fig} for Case~2.}
\label{disp_pv2av2npb0.182_128x4_256_iden0eps1e-4_dx0.75:fig}
\end{figure}

In Figures~\ref{vxvz_pv2av2_128x3_256_iden0eps1e-4_dx0.75_pa_t480-5840:fig}-\ref{disp_pv2av2npb0.182_128x4_256_iden0eps1e-4_dx0.75:fig} we show the proton and $\alpha$ particle VDFs, the temporal evolution of the temperature anisotropies, the relative drift (beam) speed of the ion core and beam populations, and the associated magnetic energy and magnetic dispersion spectrum obtained from the 2.5D hybrid model. The proton VDFs at $t=480$, 1000, 3000, and 5840 $\Omega_p^{-1}$ with initial $\beta_{p,core}=0.429$, and a cold beam with initial $\beta_{p,beam}=0.092$, and  with initial relative drift speed of $2V_A$ (Case~1) are shown in Figure~\ref{vxvz_pv2av2_128x3_256_iden0eps1e-4_dx0.75_pa_t480-5840:fig}. The upper panels show the proton VDFs in the $V_x-V_z$ phase space plane, and the lower panels show the cut of the VDFs along $V_x$ through the peak of the distributions at the corresponding times. The evolution and the structure of the non-Maxwellian core and beam VDFs are evident. The large temperature anisotropy (calculated from the ratio $T_\perp/T_\parallel$) of the beam population of protons is exhibited by the spread of the proton VDFs in $V_z$, with subsequent relaxation and merging of the core-beam proton population.  The initial super-Alfv\'{e}nic proton beam lead to the growth of magnetosonic instability, as expected from Vlasov's linear stability analysis \citep[see, e.g.,][]{Gar93,XOV04,OV07}, with subsequent nonlinear saturation and relaxation. The linear growth stage of instability takes place within the first several tens of gyroperiods. After this short time the VDF of protons (as well as $\alpha$ particles) is no longer Maxwellian as evident in anisotropy increase in Figure~\ref{aniso_pv2av2_128x3_256_iden0eps1e-4_dx0.75_12:fig} producing left-hand polarized resonant ion-cyclotron waves, evident in the dispersion relation resonant branches in Figure~\ref{disp_pv2av2_128x3_256_iden0eps1e-4_dx0.75:fig} below. The increased temperature anisotropy $T_\perp/T_\parallel>1$ of the ions leads to secondary ion-cyclotron instability driven primarily by the beam populations. The secondary instability is gradually damped due to wave-wave, wave-particle scattering, and velocity space diffusion into the parallel direction and absorption relaxation of the instability. The resulting gradual parallel heating of the ions is evident in Figure~\ref{wp_pv2av2_128x3_256_iden0eps1e-4_dx0.75_12:fig} below.

The $\alpha$ particle population VDFs at $t=480$, 1000, 3000, and 5840 $\Omega_p^{-1}$ for Case~1 with $n_{p,b}=0.092$ are shown in Figure~\ref{vxvz_pv2av2_128x3_256_iden0eps1e-4_dx0.75_pa_t480-5840:fig} (lower two panels). The $\alpha$ particle VDFs in the $V_x-V_z$ phase space plane, and the lower panels show the cut of the $\alpha$ particle VDFs along $V_x$ through the peak of the $\alpha$ particle distributions at the corresponding times. The instability leads to rapid increase of the perpendicular anisotropy of the beam ion populations. This is followed by the isotropisation of the beam on much longer time scales, where the relaxation of the beam anisotropy of the  $\alpha$ particle population proceeds more slowly compared to the proton population shown at the same time in Figure~\ref{vxvz_pv2av2_128x3_256_iden0eps1e-4_dx0.75_pa_t480-5840:fig}. Also, more gradual merging of the beam and the core $\alpha$ particle populations at the indicated times is evident.  The very large (compared to protons) temperature anisotropy  of the $\alpha$ particle beam population is exhibited by the wide spread of the $\alpha$ particle VDFs in $V_z$, with subsequent  partial relaxation of the anisotropy and near-merging of the core-beam $\alpha$ particle population at $t=5840\ \Omega_p^{-1}$.  The initial $\alpha$ particle super-Alfv\'{e}nic beam results in the magnetosonic instability in the $\alpha$ particle population that proceeds together with the proton  magnetosonic instability. 

The proton VDFs $t=240$, 480, 1000, and 3000 $\Omega_p^{-1}$ for Case~2 with the same parameters as in Case~1 but with a twice denser beam population, $n_{p,b}=0.182$ are shown in Figure~\ref{vxvz_pv2av2npb0.182_128x4_256_iden0eps1e-4_dx0.75_pa_t240-3000:fig}. The upper panels show the proton VDFs in the $V_x-V_z$ phase space plane, and the lower panels show the cut of the VDFs along $V_x$ through the peak of the distribution at the corresponding times. The more rapid (than in Case~1) growth and relaxation of the magnetosonic instability is evident.  The large temperature anisotropy of the denser beam population of protons is exhibited by the spread of the proton VDFs in $V_z$, with subsequent rapid relaxation and merging of the core-beam proton population at $t=3000\ \Omega_p^{-1}$.  Here as well, the initial proton super-Alfv\'{e}nic beam results in the magnetosonic instability in the proton population, with subsequent nonlinear saturation and relaxation that proceeds more rapidly than in Case~1 (see, Figure~\ref{aniso_pv2av2_128x3_256_iden0eps1e-4_dx0.75_12:fig} below).


The $\alpha$ particle population VDFs $t=240$, 480, 1000, and 3000 $\Omega_p^{-1}$ for Case~2 with $n_{p,b}=0.182$ are shown in Figure~\ref{vxvz_pv2av2npb0.182_128x4_256_iden0eps1e-4_dx0.75_pa_t240-3000:fig} (lower two panels). The $\alpha$ particle VDFs in the $V_x-V_z$ phase space plane is shown, and the lower panels show the cut of the $\alpha$ particle VDFs along $V_x$ through the peak of the $\alpha$ particle distributions at the corresponding times. The more rapid growth and relaxation of the magnetosonic instability in the  $\alpha$ particle population compared to the proton population shown at the same time in Figure~\ref{vxvz_pv2av2npb0.182_128x4_256_iden0eps1e-4_dx0.75_pa_t240-3000:fig} is evident.  The very large temperature anisotropy of the $\alpha$ particle beam population is exhibited by the wide spread of the $\alpha$ particle VDFs in $V_z$, with subsequent rapid relaxation and near-merging of the core-beam $\alpha$ particle population at $t=3000\ \Omega_p^{-1}$.  Here as well, the initial $\alpha$ particle super-Alfv\'{e}nic beam results in the magnetosonic instability in the $\alpha$ particle population, with subsequent nonlinear saturation that proceeds more rapidly than in the proton population.

The temporal evolution of the temperature anisotropy for the proton core and beam populations, the $\alpha$ particle core and beam populations, and the proton core-beam drift speeds, $U_{drift}$ for Cases~1 and 2 are shown for comparison in Figure~\ref{aniso_pv2av2_128x3_256_iden0eps1e-4_dx0.75_12:fig}. The more rapid growth and relaxation of the instability in Case~2 with $n_{p,b}=0.182$ than in Case~1 with $n_{p,b}=0.091$ is evident. The proton beam population temperature anisotropy in Case~1 peaks at $\sim4.5$ at $t\approx 560\ \Omega_p^{-1}$, while in Case~2 the temperature anisotropy peaks at $\sim5$ at $t\approx 230\ \Omega_p^{-1}$, with slight increase of the proton core population anisotropy. In Case~1 the $\alpha$ particle beam population reaches temperature anisotropy of $\sim5.2$, while the $\alpha$ particle core population reaches temperature anisotropy of $\sim1.8$. In Case~2 the $\alpha$ particle beam population reaches temperature anisotropy of $\sim7$ at $t=110\ \Omega_p^{-1}$, while the $\alpha$ particle core population reaches temperature anisotropy of $\sim3$. Thus, it is reasonable to conclude that the additional free energy in the proton beam produces more energetic wave spectrum that affects both, protons, and $\alpha$ particle populations. The temporal evolution of the moment of the proton parallel velocity $V_x$ of the core and beam populations shown, and the decrease of the drift speed due to the relaxation of the instability is apparent in Figures~\ref{aniso_pv2av2_128x3_256_iden0eps1e-4_dx0.75_12:fig}e-f, with more rapid evolution in Case~2 compared to Case~1.  In order to demonstrate the effects of the gradual solar wind expansion on the evolution of the beam instability we have added the ion temperature anisotropies and the proton velocity drift calculated without solar wind expansion in Case~2. It is evident that expansion leads to more rapid decrease of the ion anisotropies and the proton drift velocity, and the effect of the expansion is more significant for the $\alpha$ particle populations compared to protons. In the proton core population there is a slight decrease of the proton core anisotropy to values  $T_\perp/T_\parallel\lesssim1$ after a couple of thousand gyroperiods in Cases~1 and 2 that is followed by slight increase of anisotropy back to $T_\perp/T_\parallel\sim1$ that could be attributed to weak (slowly growing) firehose instability in the proton core population.

While it is evident that the proton-beam driven magnetosonic instability destabilizes the $\alpha$ particle population, we have evaluated the effects of the  $\alpha$ particle beam instability by comparing the combined proton and $\alpha$ beam instabilities modeled in Case~1 to the proton only beam instability (i.e., without $\alpha$ particles) in Case~3. We found that the evolution of the proton instability is very similar to the evolution of the instability in Case~1, with slightly stronger effects of the solar wind expansion on the anisotropy decrease of the core proton population in Case~3 than in Case~1. This small difference could be attributed to the secondary heating of protons due to the $\alpha$ beam instability.

The temporal evolution of the parallel kinetic energy $W_\parallel$ and perpendicular kinetic energy $W_\perp$ for the proton and $\alpha$ particle populations for Case~1 and 2 are shown in Figure~\ref{wp_pv2av2_128x3_256_iden0eps1e-4_dx0.75_12:fig}. The plots show that the parallel and perpendicular proton core energies change gradually in both cases, while the significant change takes place in the beam populations, consistent with the proton VDFs shown in Figures~\ref{vxvz_pv2av2_128x3_256_iden0eps1e-4_dx0.75_pa_t480-5840:fig}-\ref{vxvz_pv2av2npb0.182_128x4_256_iden0eps1e-4_dx0.75_pa_t240-3000:fig}. The $W_\parallel$ of the proton beam population initially decreases, resulting in increased temperature anisotropy, shown in Figure~\ref{aniso_pv2av2_128x3_256_iden0eps1e-4_dx0.75_12:fig}a-b, with stronger effect in Case~2 compared to Case~1. The $W_{p,\parallel}$ decreases gradually in both cases with slight initial decrease of $W_{p,\parallel}$ in the proton beam population in Case~2, and corresponding increase of the  $W_{p,\perp}$, followed by further gradual decrease as expected due to the gradual SW expansion. The $\alpha$ particle core $W_{\alpha,\perp}$ shows rapid increase due to the onset of the magnetosonic instability in Case~1 and 2 with more rapid increase in Case~2 compared to Case~1, consistent with the evolution of the $\alpha$ particle temperature anisotropy shown in Figure~\ref{aniso_pv2av2_128x3_256_iden0eps1e-4_dx0.75_12:fig}c-d.

In Figure~\ref{2dfft_kxky:fig} we show the $k$-plane power spectra of the perpendicular magnetic fluctuations due to the beam  instabilities modeled in Cases~1 and 2 with the 2.5D hybrid code. In Case~1 it is evident that initially at the growth phase of the instability ($t=240\Omega_p^{-1}$) wave power is large in oblique modes at $k_x\approx0.45\delta_p^{-1}$, $k_y\approx0.4\delta_p^{-1}$ as well as in the parallel mode consistent with the linear Vlasov's stability analysis at $k_x\approx0.7\delta_p^{-1}$. The oblique modes damp, and the remaining wave power is in the unstable parallel mode, evident at $t=480\Omega_p^{-1}$. At later time, $t=1000\Omega_p^{-1}$ the unstable mode has damped, and only longer wavelength modes remain in the dissipation phase of the beam instability. Qualitatively similar evolution is evident in the more unstable Case~2, where at early time $t=120\Omega_p^{-1}$ the wave power is peaked  near $k_x\approx0.5\delta_p^{-1}$ with significant power in the oblique modes $|k_y|\lesssim0.5$. At the later time $t=240\Omega_p^{-1}$ the oblique modes damp, and at  $t=480\Omega_p^{-1}$ the remaining power is concentrated at the parallel mode, peaking at $k_x\approx0.35\delta_p^{-1}$. It is evident that the power is concentrated in short wavelength wave, with typical $k~0.35-0.7\delta_p^{-1}$ corresponding the wavelengths in the range $9-18\delta_p$. With the PSP plasma parameters measured at E4 these values would correspond to wavelengths of 65-135 km in the solar wind rest frame.

The dispersion relation calculated from the perpendicular magnetic fluctuations $B_\perp$ from the 2.5D hybrid model with the beam parameters of Case~1 is shown in Figure~\ref{disp_pv2av2_128x3_256_iden0eps1e-4_dx0.75:fig} (left panel). The various branches of the dispersion relation and the fastest growing modes identified as dark blue are evident. Comparing to the dispersion relations calculated from the 2.5D hybrid code \citep[e.g.,][]{XOV04,OV07,OOV15,OVR17}, we can identify the resonant modes that show the largest intensities at $|k_\parallel|\sim0.3$ and strong damping at $|k_\parallel|\gtrsim 1$, and the right-hand polarized non-resonant modes of weaker intensity appearing undamped at higher $\omega$ and $|k_\parallel|$ crossing and exceeding the proton gyroresonat frequencies $\omega=\pm\Omega_p$. The magnitude of the slope of the resonant branches is approximately $V_A$, as expected from the asymptotic stability threshold beam speed and their branches are shifted by about $0.5\Omega_p$ from the right-hand modes, indicating that these branches are produced by the $\alpha$ particle beam Doppler shifted with respect to the core population. It is interesting to note that the linear Vlasov's dispersion (Figure~\ref{displin:fig}) predicts higher $|k_\parallel|\sim0.76$ for the fastest growing mode for the proton parameters of Case~1 and for $\alpha$ particles the peak growth is expected for half of this wavelength due to their smaller inertial length with respect to protons, i.e., $\delta_\alpha/\delta_p=(m_\alpha/m_p)^{-1/2}=0.5$. Consistently, it is evident from the dispersion relation obtained with the hybrid model that the highest power of the resonant branches is in the range $|k_\parallel|\sim0.3-0.7$. 

The drifting population of the ions produces the Doppler shifted cyclotron resonances $\omega_i-k_\parallel V_\parallel=\Omega_i$ \citep[e.g.][]{Sti62}. While, both, the proton, and the $\alpha$ particle beam population resonances are shifted to higher frequencies,  the $\alpha$ population may resonate with the $\alpha$ and proton-produced beam instability wave spectrum. Moreover, the $\alpha$ particle core population can affect the ion-cyclotron wave dispersion by introducing an absorption gap in the parallel-propagating waves \citep[e.g.][]{CS71,COD88}.  Previous studies have considered proton-$\alpha$ population and ion population drift, rather than the proton-proton and $\alpha-\alpha$ beams of the present study, there are similarities in the dispersion wave spectra with the additional effects of the Doppler shift \citep[see, e.g.][]{XOV04,OVR17}. The average perpendicular magnetic energy $\left<|B_\perp|^2\right>$  shown in the right panel increase rapidly in time with the growth of the magnetosonic instability, following saturation at nearly constant level, consistent with the evolution of the instability and the ion temperature anisotropies for Case~1 shown in Figure~\ref{aniso_pv2av2_128x3_256_iden0eps1e-4_dx0.75_12:fig}a and c. The initial peak in $\left<|B_\perp|^2\right>$ can be associated with the peak of the $\alpha$ beam temperature anisotropy evident in Figure~\ref{aniso_pv2av2_128x3_256_iden0eps1e-4_dx0.75_12:fig}d, and the relaxation of the $\alpha$ population beam-driven instability.

The dispersion relation calculated from the perpendicular magnetic fluctuations $B_\perp$ from the 2.5D hybrid model with the beam parameters of Case~2 is shown in Figure~\ref{disp_pv2av2npb0.182_128x4_256_iden0eps1e-4_dx0.75:fig} (left panel). Here as well, the  branches of the dispersion relation and the fastest growing modes identified as dark blue are evident. The more energetic, faster growing instability in Case~2 results in stronger wave activity compared to Case~1 as evident from the more extensive high intensity (blue)  regions in the dispersion relation branches showing evidence of both, $\alpha$ and weaker proton branches. Comparing the dispersion relation to Case~1 and the previous studies mentioned above, we can identify the resonant modes that show the largest intensity in the range $|k_\parallel|\sim0.1-0.8$ and strong damping for $|k_\parallel|>1$, while the non-resonant modes of weaker intensity proceed undamped at higher $\omega$ exceeding the proton gyroresonat frequency.  The $\alpha$ particle branches are evident with an approximate slope magnitude of $V_A$, the asymptotic stability threshold, and shifted by about $0.5\Omega_p$ with respect to the non-resonant branches, with apparent Doppler-shifted proton branches at higher frequencies. However, the combined effects of power obtained from the nonlinear evolution of the ion resonances on the dispersion spectrum is not well separable in the present case.  The total perpendicular magnetic energy increases rapidly with the growth of the magnetosonic instability within $\sim110\omega_p^{-1}$, following rapid decrease and saturation at nearly constant level, consistent with the evolution of the instability and the ion temperature anisotropies for Case~2 shown in Figure~\ref{aniso_pv2av2_128x3_256_iden0eps1e-4_dx0.75_12:fig}b and d.

\subsection{3D hybrid modeling results}
\begin{figure}[h]
\centering
\includegraphics[width=0.9\linewidth]{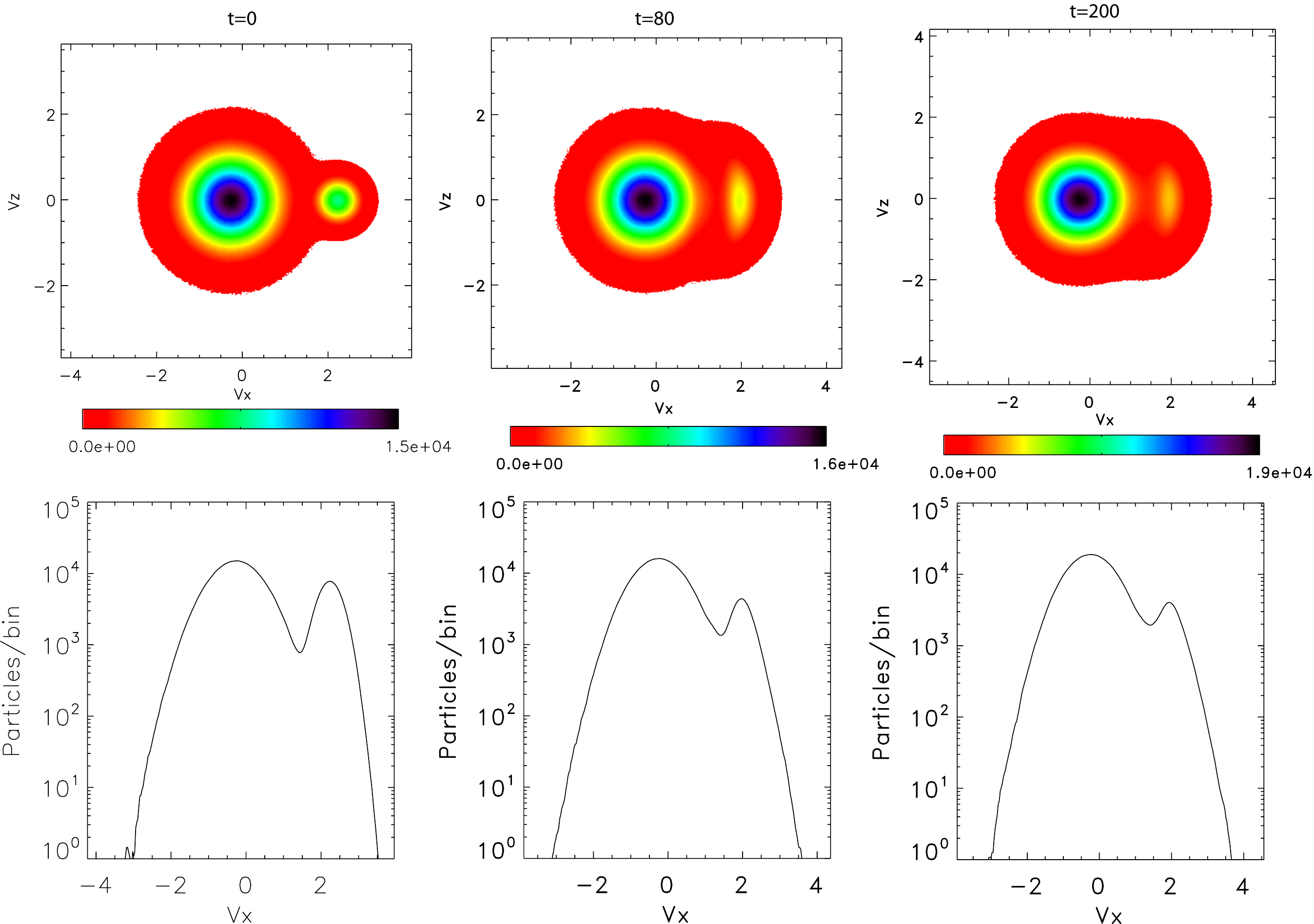}
\caption{The VDFs obtained from the 3D hybrid model for Case~4 with $V_d=2.5V_A$. The top panels show the proton VDFs in the $V_x-V_z$ phase space plane in the initial state (t=0) and at $t=80$, and 200 $\Omega_p^{-1}$. The lower panels show the cut along $V_x$ through the peak of the VDFs.}
\label{vxvz_pv2.5_128_32x2_b0.429bb0.0916_t0-200:fig}
\end{figure}

\begin{figure}[h]
\centering
\includegraphics[width=0.5\linewidth]{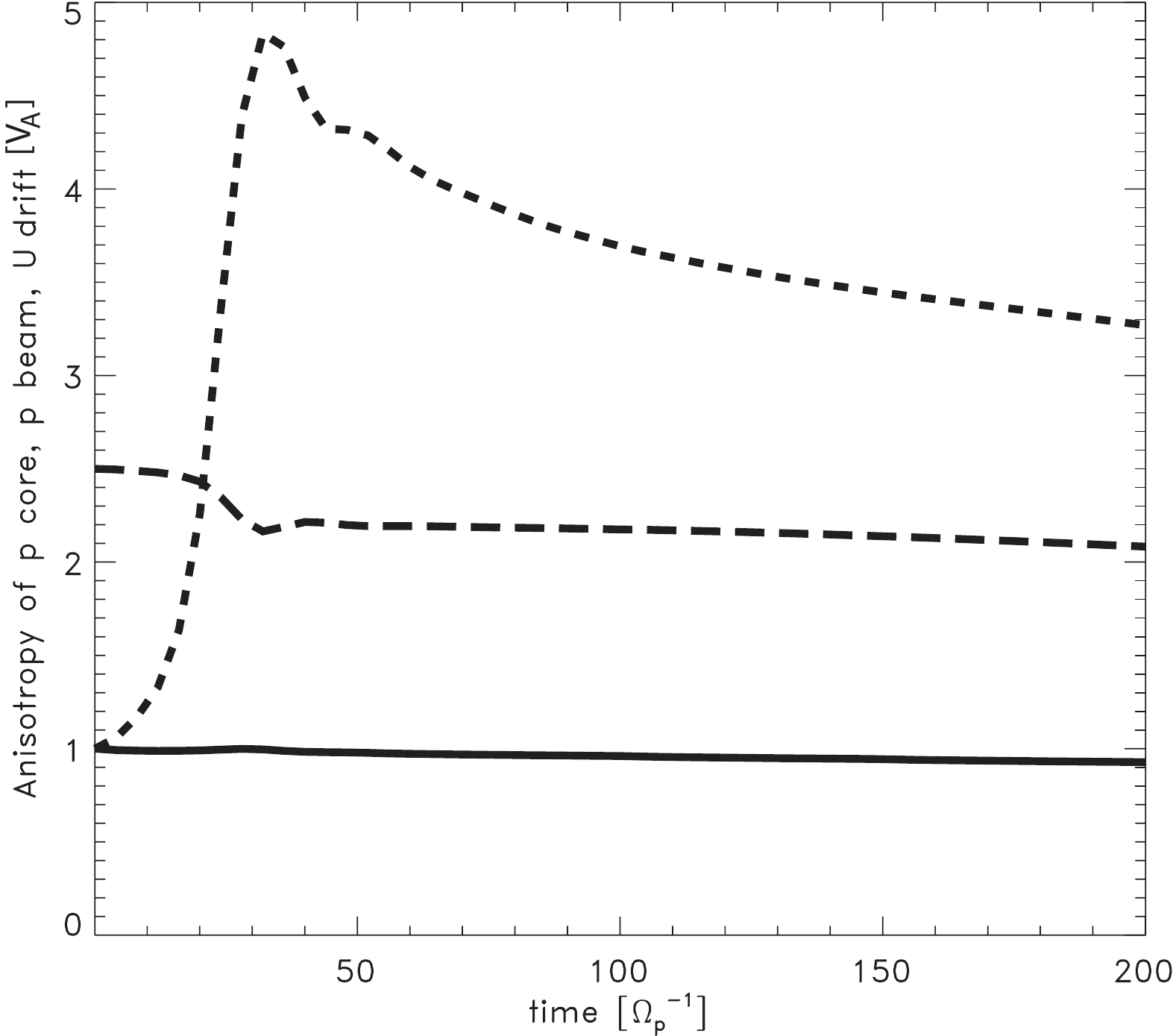}
\caption{The temporal evolution of the proton core population temperature anisotropy (solid), proton beam population temperature anisotropy (short dashes) and the core-beam drift speed $U_{drift}$ (long dashes) obtained from the 3D hybrid model for Case~4 with $V_d=2.5V_A$. }
\label{aniso_3d_pv2.5_128_32x2_b0.429bb0.0916_t200:fig}
\end{figure}

Figures~\ref{vxvz_pv2.5_128_32x2_b0.429bb0.0916_t0-200:fig}-\ref{aniso_3d_pv2.5_128_32x2_b0.429bb0.0916_t200:fig} show the results of the 3D hybrid modeling for Case~4 with an initial proton beam speed of 2.5$V_A$ with respect to the core. The number density of the beam is 10\% of the total proton population, with 90\% of the protons in the core. It is evident in Figure~\ref{vxvz_pv2.5_128_32x2_b0.429bb0.0916_t0-200:fig} that the initial Maxwellian VDF of the beam becomes anisotropic with increased $T_{p,\perp}$, marked by the spread of the beam VDF in the perpendicular ($V_z$) direction at $t=80$, and 200 $\Omega_p^{-1}$. The proton core VDF remains nearly Maxwellian at this relatively short timescale (compared to the 2.5D runs in Cases~1-3).

The temporal evolution of the proton core and beam populations temperature anisotropies, and the proton beam-core drift speed $U_{drift}$ for Case~4 is shown in Figure~\ref{aniso_3d_pv2.5_128_32x2_b0.429bb0.0916_t200:fig}.  The initial growth of the beam-driven magnetosonic instability results in increase of the temperature anisotropy to $4.3$ within $\sim32$ $\Omega_p^{-1}$, with subsequent decrease to $3.3$ at $t=200\ \Omega_p^{-1}$. The corresponding initially rapid decrease of the beam-core drift speed is evident, followed by gradual decrease. The core proton population remains nearly isotropic throughout the modeled evolution to $t=200\ \Omega_p^{-1}$, with very gradual decrease, with contribution to the decrease by the solar wind expansion, where the expansion parameter was set at $\epsilon=10^{-4}$ in the 3D hybrid model (same as used in the 2.5D hybrid model).

In order to evaluate the importance of the 3D effects on the evolution of the proton beam-driven magnetosonic  instability, we have compared the 3D hybrid model results to the results of 2.5D hybrid model with otherwise identical physical and numerical parameters. We found that the temporal evolution of the temperature anisotropies and relaxation rate of the instability in the 3D model is close to the 2.5D model, and the results of the two models are similar. The main difference is slightly faster onset and initial growth of the instability in 2.5D hybrid model compared to the 3D hybrid model result, where the peak temperature anisotropy of the beam is reached $\sim10\ \Omega_p^{-1}$ sooner than in 3D hybrid model, but, reaching close maximal proton beam temperature anisotropy of $\sim4$ in the two models. 

\section{Discussion and Conclusions} \label{disc:sec}
Recent PSP SPAN-I observations near perihelia find evidence of proton and $\alpha$ particles non-Maxwellian VDFs with core and beam populations, with FIELDS instrument detecting related periods of high ion-scale wave activity in the solar wind plasma at PSP perihelia (Encounter 4) with further observations of proton beams in Encounters 4-8 \citep{Ver20,Ver21}. Motivated by PSP observations, we use 2.5D and 3D expanding box hybrid modeling to investigate the growth and nonlinear state of the instabilities and kinetic waves produced by proton and $\alpha$ beams, and a temperature anisotropy using hybrid simulations. The parameters of the models such as ion $\beta_\parallel$ and the core-beam super-Alfv\'{e}nic differential speed are guided by the PSP ion data. We find good qualitative agreement between the modeled VDFs and the proton PSP SPAN-I observations, that show similar core-beam bump-on-tail structure for protons and $\alpha$ particles. The modeled dispersion of the ion-scale waves and the $k$-spectrum evolution show strong power in ion scale waves identified in the  model as resonant ion-cyclotron and non-resonant magnetosonic waves. The PSP FIELD instrument has provided evidence of periods of ion-scale wave activity with left-hand and right-hand polarization, in qualitative agreement with the model results. However, direct comparison to the PSP/FIELDS observations obtained in the spacecraft frame to the  results in the solar wind frame that model the evolution of a particular plasma parcel, are ambiguous due to the effects of  Doppler shift, and the possible variation of the solar wind plasma properties  (possibly unrelated to the evolution of the beam instability) as it flows by the spacecraft detectors. The hybrid models show that super-Alfv\'{e}nic beams lead to rapid (in terms of $\Omega_p^{-1}$) growth of the magnetosonic instability and in agreement  with Vlasov's linear theory, reaching quickly the nonlinear saturated state, with significant temperature anisotropy increase of the proton and $\alpha$ particle beam populations (peaking at $\sim4$ for the modeled parameters), and with smaller anisotropy of the core populations, followed by gradual relaxation of the drift and anisotropies.  We note that the peak anisotropies for the proton beam in Case 1 of the model exceed 4 for about 400 gyroperiods ($\sim4$ min) and for 250 gyroperiods ($\sim$2.5 min) in Case~2, with most of the modeled evolution time of several thousand gyroperiods in agreement with the range of anisotropies obtained from PSP observations. The PSP data in Figure~\ref{psp_fields01262020:fig} show anisotropies as high as ~4 at t=13:50-13:52 and 14:44-14:45 comparable to the peak anisotropies in the model.

We find that  the instability growth rate and nonlinear saturation level of the temperature anisotropy and magnetic wave energy increase with the proton beam population density. The initial growth stage takes place within several tens of proton gyroperiods as predicted by linear theory, and relaxation of the instability occurs over several thousand gyroperiods with the beam protons eventually merging with the core proton populations through phase-space diffusion, combined with a small effect of perpendicular temperature decrease due to SW expansion, in agreement with previous studies. The $\alpha$ particle population shows faster growth of the instability for the same beam speed and more gradual relaxation than protons of the beam-driven instability. This is  due to the combined effects of the larger $\alpha$ beam-to-core density ratio and the absorption of some of the proton-beam produced wave-spectrum energy Doppler shifted to the resonant frequency range, as evident in the dispersion relation. While the interaction between the proton and $\alpha$ beams is not strong due to their separation in the wave dispersion, some of the proton instability produced wave spectrum is absorbed by the $\alpha$ particles due to the Doppler shift of the beam populations. The perpendicular heating of the  ions leads to a secondary instability driven by their temperature anisotropy.  We extend the hybrid model of the proton-beam driven instability to the full 3D expanding box model and find similar results as in the 2.5D hybrid model for the same physical parameters, indicating that the main effects  of the growth and saturation of the instability are due to the parallel and planar propagating waves. Both the 2.5D and 3D hybrid models produce the 3D VDFs of protons and $\alpha$ particles that are in qualitatively good agreement PSP SPAN-I observations.

The LJ and LO acknowledge support by NASA LWS grant 80NSSC20K0648. LO acknowledges support by NASA Cooperative Agreement NNG11PL10A to The Catholic University of America, and NASA Partnership for Heliophysics and Space Environment Research (PHaSER) award 80NSSC21M0180.  JV and DL acknowledge the support by NASA contract NNN06AA01C. We thank Michael McManus for the help in producing Figures 1 and 2. We thank the PSP mission team for generating the data and making them publicly available. Resources supporting this work were provided by the NASA High-End Computing (HEC) Program through the NASA Advanced Supercomputing (NAS) Division at Ames Research Center. The authors thank Dr. Vadim Roytershtein for providing the linear Vlasov dispersion solver.

%

\vspace{5mm}
\facilities{PSP (SWEAP, FIELDS)}

\end{document}